\documentclass[reprint, aps,superscriptaddress,amsmath,amssym,prl]{revtex4-2}

\usepackage{bm}
\usepackage{float}
\usepackage{graphicx}
\usepackage[usenames,dvipsnames]{color}
\usepackage[normalem]{ulem}
\usepackage[svgnames]{xcolor}
\usepackage{bm}
\usepackage{multirow}
\usepackage{titlesec}
\usepackage[utf8]{inputenc}
\usepackage[colorlinks,linkcolor=blue,citecolor=blue,urlcolor=blue]{hyperref}
\usepackage{booktabs}
\usepackage{array}
\usepackage{url}
\usepackage{tabularray}

\begin{document}
\title{Hydrodynamics of the viscous electron fluid in cadmium}

\author{Xiaodong Guo}
\affiliation{Wuhan National High Magnetic Field Center and School of Physics, Huazhong University of Science and Technology,  Wuhan  430074, China}
\affiliation{Laboratoire de Physique et d'\'Etude de Mat\'{e}riaux (CNRS)\\ ESPCI Paris, PSL Research University, 75005 Paris, France }

\author{Xiaokang Li}
\affiliation{Wuhan National High Magnetic Field Center and School of Physics, Huazhong University of Science and Technology,  Wuhan  430074, China}

\author{Beno\^it Fauqu\'e}
\affiliation{JEIP, USR 3573 CNRS, Coll\`ege de France, PSL University, 11, place Marcelin Berthelot, 75231 Paris Cedex 05, France}

\author{Alaska Subedi} 
\affiliation{CPHT, CNRS, \'Ecole polytechnique, Institut Polytechnique
  de Paris, 91120 Palaiseau, France} 

\author{Lingxiao Zhao}
\email{zhaolingxiao@quantumsc.cn}
\affiliation{Quantum Science Center of Guangdong-HongKong-Macao Greater Bay Area, Shenzhen 523335, China}
\affiliation{Department of Physics, Southern University of Science and Technology, Shenzhen, China}

\author{Zengwei Zhu}
\email{zengwei.zhu@hust.edu.cn}
\affiliation{Wuhan National High Magnetic Field Center and School of Physics, Huazhong University of Science and Technology,  Wuhan  430074, China}

\author{Kamran Behnia}
\email{kamran.behnia@espci.fr}
\affiliation{Laboratoire de Physique et d'\'Etude de Mat\'{e}riaux (CNRS)\\ ESPCI Paris, PSL Research University, 75005 Paris, France }

\begin{abstract}
Thanks to electron-electron ($e$-$e$) collisions  conserving momentum,  metallic electron fluids are viscous. Yet, this viscosity is rarely detectable in bulk transport. Here,  we report on the canonical realization of the Gurzhi effect in an elemental three-dimensional metal: cadmium. Using focused ion beam microstructuring to tune the effective thickness, we detected a low-temperature size-dependent resistivity upturn in a finite window sandwiched between ballistic and diffusive regimes. Within this window, the electrical conductivity displays a simultaneous quadratic dependence on both sample size and temperature— fingerprint of a hydrodynamic flow. This leads us to quantify the amplitude and the temperature dependence of kinematic and dynamic viscosity of the electron fluid. In cadmium, in contrast with graphene and $^3$He, the rate of momentum-conserving $e$-$e$ collisions is not set by the main Fermi energy, but by Lilliputian energy scales and inter-valley bottlenecks.
\end{abstract}
\maketitle

According to empirical knowledge, accessible in a kitchen, warming a liquid reduces its viscosity. This is also true for quantum liquids such as normal $^3$He at cryogenic temperatures: warming enhances fermion-fermion collisions rate, diminishing viscosity \cite{Dobbs}.  In apparent contrast,  the electrical conductivity of a metallic electronic liquid worsens by warming. The fundamental reason for this difference is that most collisions suffered by electrons, instead of conserving their momentum, relax it. Their multiplication enhances resistivity and not conductivity. 

However, hydrodynamic corrections to electronic transport are expected to emerge whenever the quasi-particle momentum is approximately conserved \cite{Polini-viscouselectronliquid,Fritz2024,hui2025}. Decades ago, Gurzhi \cite{gurzhi1968,gurzhi-2} assumed a metallic system where the momentum-conserving (MC) mean free path, $\ell_{mc}$, becomes  shorter than both the sample size  and the momentum-relaxing (MR) mean free path, $\ell_{mr}$. In this case, the momentum and energy of the traveling electrons is no longer lost to the outside bath. The flow becomes viscous, and warming would reduce the electric resistivity of the metal \cite{Polini-viscouselectronliquid}.

This appealing picture is hard to realize experimentally. For $\ell_{mc}$ to become the shortest length scale, the temperature should be sufficiently low (to make collision by phonons negligible) yet finite (to have sufficiently frequent collisions between electrons). The phase space for $e$-$e$ collisions follows $(T/T_F)^2$. Most of these collisions, presumably because they are Umklapp events, do not conserve momentum. The experimental fingerprint of $e$-$e$ collisions  has been a $T$-square enhancement of electrical resistivity (and not conductivity) \cite{Behnia2022}.

\begin{figure*}[hbt!]
\centering
\includegraphics[width=1.0\linewidth]{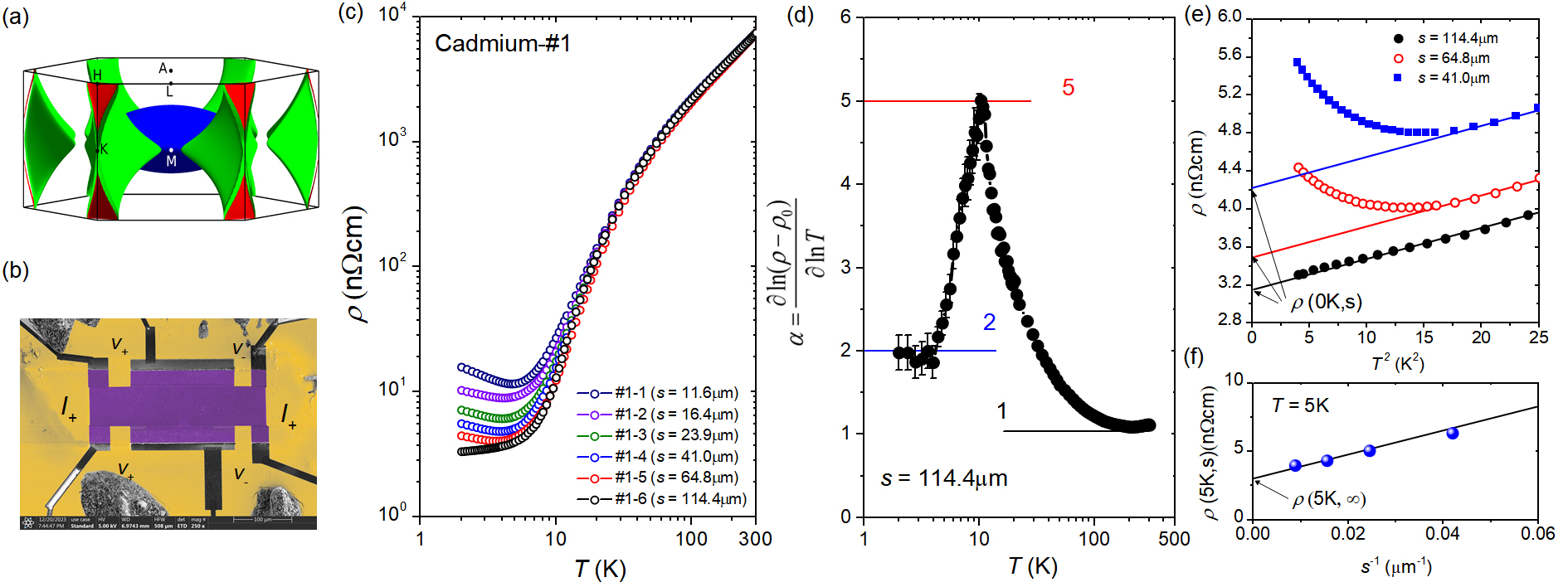} 
\caption{\textbf{Size dependence of electrical resistivity.} \textbf{(a)} Fermi surface of HCP Cd consisting of hole-type ‘cap’ (red), hole-type ‘monster’ (green), and electron-type ‘lens’ (blue) sheets. \textbf{(b)} False color image of Cd microstructures (purple) and electrodes (yellow) obtained with scanning electron microscope. In this lamella, the sample thickness is oriented along the $[0001]$ direction, and the current $I$ is applied along the$[\bar{1}2\bar{1}0]$ direction. \textbf{(c)} The evolution of the temperature dependence of electrical resistivity of a cadmium crystal with effective thickness ($s = \sqrt{w \times t}$), the geometric average of width ($w$) and thickness ($t$). Below 5 K, an upturn emerges. The thickness of samples $\text{\#1-1}$ to $\text{\#1-5}$ was kept constant while $w$ was progressively reduced. \textbf{(d)} The exponent of the inelastic resistivity ($\rho=\rho_{0}+A T^\alpha$), determined from the logarithmic derivative of $(\rho-\rho_{0})$ in sample $s=114.4~\mu$m as a function of the temperature. \textbf{(e)} Resistivity as a function of the square of temperature. The slope of the linear fit ( $A\simeq 0.03 \pm 0.002$ $\rm{n\Omega cm K^{-2}}$ ) does not change with $s$. The intercept increases with decreasing $s$.  \textbf{(f)} Resistivity at 5 K  as a function of the inverse of $s$. As expected in a ballistic picture, the dependence is linear with a slope of $a \simeq 8.8\pm 0.3 \times 10^{-16}$ $\rm{\Omega \cdot m^2}$.}
\label{fig.1}
\end{figure*}
In the past few years, experiments have documented a hydrodynamic regime sandwiched between diffusive and ballistic regimes in several metals. Imaging techniques have found that in this regime, the electron drift velocity is attenuated at the boundaries, due to the viscosity-induced shear forces at the channel walls, a phenomenon dubbed Poiseuille flow in the classical fluid mechanics  \cite{graphitedirect,graphitevisualizing,imagingscience.abm6073, obsscience.adj2167}. Corrections to transport properties due to hydrodynamics have also been reported \cite{moll2016evidence,van2021sondheimer,graphitechannel,PhysRevLettnegative,Hallscience.aau0685} notably in thermal transport \cite{Violation-wfl,gooth2018-quasi2,jaoui2021thermal}. With the exception of 2D graphene \cite{gra-2016,Polini-viscouselectronliquid,graphitechannel}, however, a clear bulk hydrodynamic behavior, in which $e$-$e$ collisions enhance conduction (instead of resistance) has remained elusive.  

In this context, Focused Ion Beam (FIB) etching technology \cite{moll2018,guo2026scalable} has emerged as an efficient tool for shrinking without introducing additional defects. By reducing the sample width while keeping $\ell_{mr}$ under control, one can fine tune the width of the hydrodynamic window sandwiched between the diffusive and ballistic transport regimes \cite{moll2016evidence,gooth2018-quasi2}. 

Here, we report on the observation of hydrodynamic electron flow that constitutes the canonical realization of the Gurzhi effect in micron-sized cadmium samples, tailored by FIB from high purity single crystals with a residual resistivity of $\rho_0$ $\sim$ 3 n$\Omega$cm and a residual resistivity ratio (RRR$=\rho(300\rm{K})/\rho_0$) of $\sim$ 3000. We find that as the effective thickness shrinks from 114.4 $\mu$m to 11.6 $\mu$m, transport at cryogenic temperatures evolves from diffusive to hydrodynamic. A canonical hydrodynamic behavior, in which conductivity scales with the square of temperature and the square of the size, is observed for the first time in an elemental three-dimensional metal. As expected, the hydrodynamic regime is restricted to a finite window of temperature and thickness. At very low temperatures and in very thin samples, a ballistic regime is recovered. We quantify the dynamic and kinematic viscosity of this electron liquid and compare it to those of classical and quantum fluids. The rate of momentum-conserving $e$-$e$ collisions in cadmium is orders of magnitude smaller than the momentum-relaxing ones, which is set by the large Fermi energy of the main pockets.  Our \textit{ab initio} DFT calculations of the electronic band structure identify narrow bottlenecks between pockets and valleys of the Fermi surface with low energy scales allowing frequent momentum exchange in small fermion reservoirs and across valleys and pockets. 

Hexagonal close-packed (HCP) cadmium is a compensated metal \cite{Subedi2024,guo2026scalable}. Its Fermi surface (Fig.\ref{fig.1}\textbf{a}) consists of an electron pocket (the `lens') and two hole pockets (the `cap' and the `monster'). We tailored samples of mesoscopic dimensions with electrodes to measure their resistivity (Fig.\ref{fig.1}\textbf{b}) . The evolution of the electrical resistivity with temperature and size is shown in Fig.\ref{fig.1}\textbf{c}. At room temperature, $\rho(273 \rm{~K})\simeq$ 6.5$~\mu\Omega$cm, consistent with what has been reported in bulk cadmium \cite{hall1970survey}. This consistency  continues down to 20 K and below this temperature a size dependence emerges. The largest sample had an effective cross-sectional area of $s=114.4~\mu$m ($s=\sqrt{w\times t}$, $w$ is the width  and $t$ the thickness) displays the metallic behavior of bulk cadmium. Those with $s \leq 64.8~\mu$m,  display a non-monotonic resistivity, which is steadily amplified with decreasing $s$ decreases (see the supplementary Note 1 \cite{SM} for details on sample dimensions).  In the thinnest sample ($s=11.6~\mu$m), at $T=$ 2 K, resistivity becomes 1.4 larger than its minimum value at 4.6 K. Absent Anderson localization or Kondo impurities, the observed resistivity upturn observed can be safely identified as the one expected when the electron fluid flows viscously. It is reminiscent of graphene constrictions, in which for a width of 0.5 $\mu$m and a density of $n = 0.5\times 10^{12} {\rm{~cm}}^{-2}$ resistivity at 2 K rises to a value $\sim 1.2$ times its minimum value at 150 K \cite{Polini-viscouselectronliquid,gra-2016,graphitechannel}.

\begin{figure*}[ht]
\centering
\includegraphics[width=0.95\linewidth]{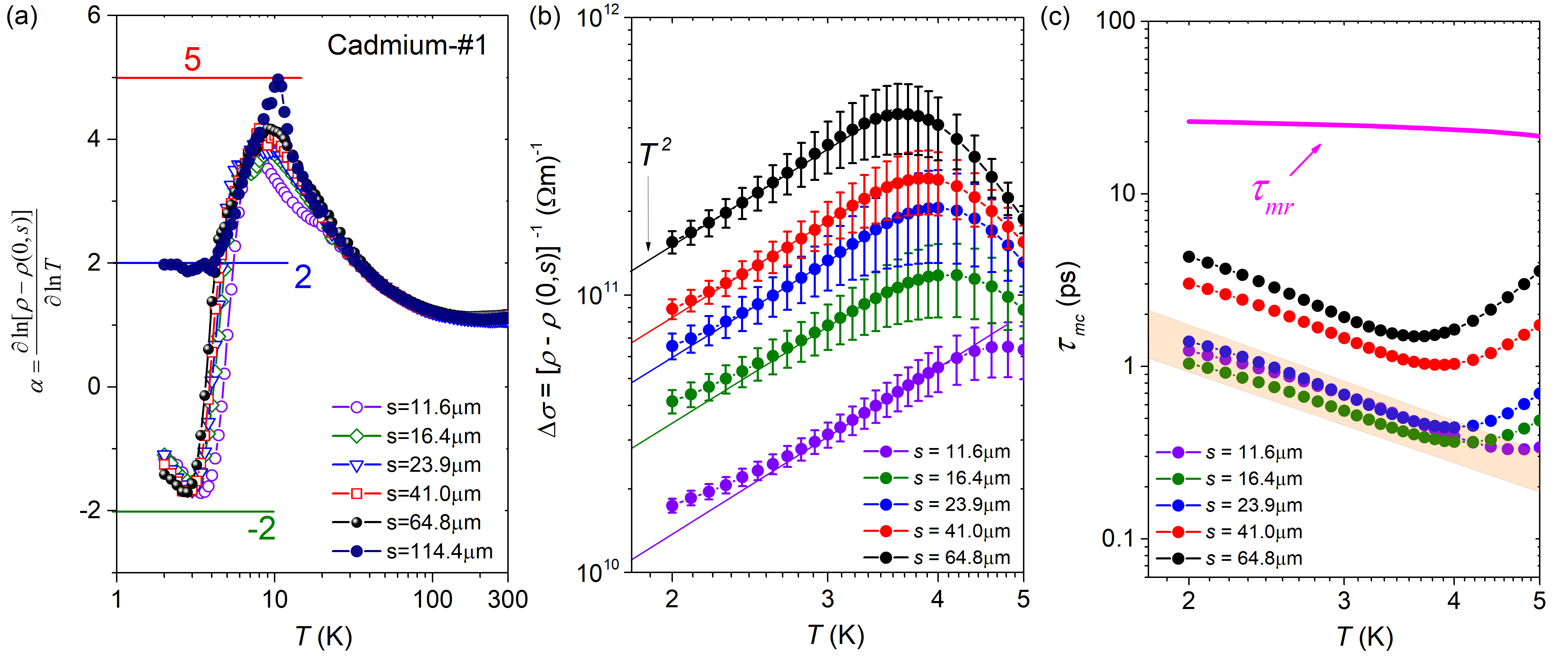} 
\caption{\textbf{Extracting the hydrodynamic contribution to electrical conductance.} (\textbf{a}) Exponent of the temperature dependence of the electrical resistivity, $\alpha$ as function of the temperature determined by taking the logarithmic derivative after subtracting $\rho (0, s)= \rho (0, \infty) + as^{-1}$. Below 5 K,  $\alpha$ approaches -2, indicating the emergence of a conductivity quadratic in temperature. (\textbf{b}) Temperature dependence of $\Delta \sigma =1/[\rho-\rho(0,s)]$. In low temperature, $\Delta \sigma \propto T^{2}$.  (\textbf{c}) Momentum-conserving scattering time, $\tau_{mc}$,  extracted from $\Delta \sigma$  and Eq.\ref{delta-s} in different samples. Also shown is the momentum-relaxing scattering time, $\tau_{mr}$. In all samples, the hydrodynamic hierarchy holds: $\tau_{mc} \ll \tau_{mr}$. $\tau_{mc}$ becomes similar in the thinnest samples.}
\label{fig.2}
\end{figure*}

The standard description of electrical transport in elemental metals is built on Bloch-Gr\"{u}neisen (BG) picture \cite{ziman2001electrons}.  Electron-phonon scattering leads to $\rho \propto$ $T$ at high temperatures and to $\rho \propto$ $T^5$ at low temperatures (when the typical phonon wave-vector is much smaller than the Fermi radius). The relevance of this picture can be checked by expressing resistivity as $\rho = \rho_0 + AT^{\alpha}$ and extracting $\alpha$ from the experimental data. Taking a logarithmic derivative leads us to $\alpha$ = $\frac{\partial \rm{ln}(\rho-\rho_0)}{\partial {\rm{ln}}T}$. The result of this analysis for the  thickest sample ($s=114.4\rm{~\mu}$m) is shown in Fig.\ref{fig.1}d. As previously noticed for other semi-metals \cite{jaoui2022formation},  $\alpha$ steadily increases upon cooling: At room temperature, $\alpha \simeq 1 $ and  at 10 K, $\alpha \simeq 5$. Below this temperature, however, $\alpha$ suddenly drops. Below 5 K, $\alpha = 2 $. This purely quadratic temperature dependence corresponds to a regime in which the only channel of inelastic scattering is resistive collisions with other electrons \cite{jaoui2022formation}. 

Fig.\ref{fig.1}\textbf{e} plots the low temperature resistivity as a function of $T^2$ in the three thickest samples. The diffusive picture of charge transport is disrupted by electron hydrodynamics below a threshold sample size.  In the thicker sample ($s = 114.4~\mu$m), resistivity is the sum of constant ($\rho_0=3.25$ n$\Omega $$\cdot$cm) and  quadratic ($AT^2$). The prefactor of the latter ($A=0.03$ n$\Omega$$\cdot$cm$\cdot$K$^{-2}$) fills in the trend seen across elemental semi-metals \cite{Gourgout2024} (See supplementary Note 4 \cite{SM}). In thinner samples, a low-temperature upturn emerges on top of this $T-$quadratic resistivity. Fig.\ref{fig.1}\textbf{e} shows that shrinking $s$ modifies the intercept ($\rho_0$), but not the slope ($A$). To quantify $\alpha$ in samples with resistivity upturn, the rigid upward shift of finite-temperature resistivity induced by thinning is to be taken into account.  

Fig.\ref{fig.1}f is a plot of 5 K resistivity as a function of the inverse of the effective thickness. The reference was chosen to be 5 K because hydrodynamic effects are absent and the exponent of resistivity is close to 2. One can see that $\rho(5  {\rm{~K}}, s)$ increases linearly with  $s^{-1}$, as expected in the diffusive-ballistic picture. The 5 K resistivity of a sample with effective thickness of $s$ is equal to  $\rho(5  {\rm{~K}},s) = \rho(5{\rm{~K}}, \infty)+ as^{-1}$. The slope of this linear enhancement ($a \simeq 8.8 \pm 0.3 \times 10^{-16} \rm{~\Omega \cdot m^2}$), set by boundary scattering, is in quantitative agreement with the computed electron density \cite{Subedi2024} and the Fuchs-Sondheimer \cite{Zhou2018} ballistic transport picture (See supplementary note 2 \cite{SM}). The intercept of the linear fit in  Fig.\ref{fig.1}d ($\rho(5 \rm{~K}, \infty)=$ 3 $\rm{n\Omega\cdot cm}$) represents 5 K resistivity in an infinite sample with a disorder level equal to samples of finite size. Since varying thickness does not affect $A$, the zero-temperature resistivity of an infinite sample is: $\rho(0 {\rm{~K}}, \infty)= \rho(5 {\rm{~K}},\infty )- 5^2A$ = $2.3\pm 0.05$ $\rm{n\Omega \cdot cm}$. This is the residual resistivity due to scattering by defects (and not the boundary). 
 
 \begin{figure*}[ht]
\includegraphics[width=1.0\linewidth]{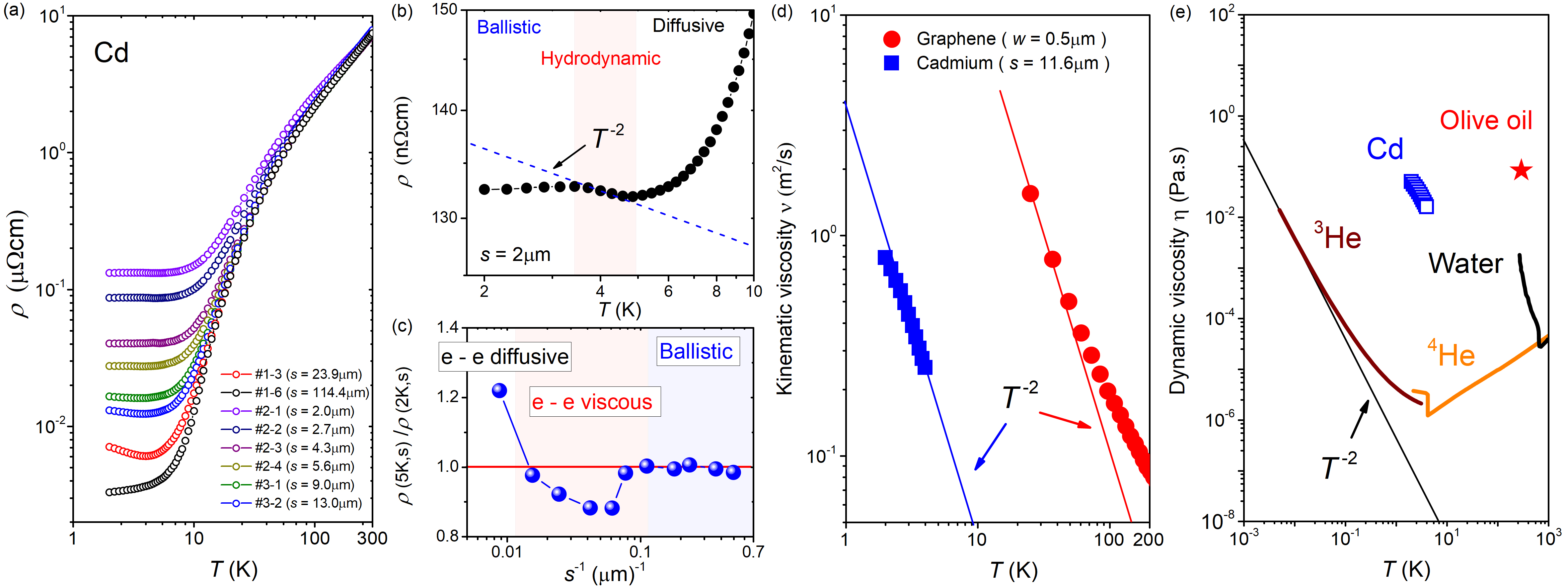} 
\centering
\caption{\textbf{Hydrodynamic to ballistic transition and viscosity comparison.} (\textbf{a}) Variation of resistivity with decreasing $s$ from $114.4~\rm{\mu m}$ to $2~\rm{\mu m}$. The mechanism turns from diffusive to hydrodynamic and then to ballistic. (\textbf{b}) The temperature dependence resistivity for $s = 2~\rm{\mu m}$ sample. Resistivity increases between 5~K to 3.5~K, and then remains constant, indicating a transition from hydrodynamic to ballistic transport. (\textbf{c}) The ratio of $\frac{\rho({\rm{5 K}},s)}{\rho({\rm{2 K}},s)}$ as a function of $s^{-1}$. (\textbf{d}) Kinematic viscosity  as function of temperature in Cd and in graphene \cite{Polini-viscouselectronliquid,graphitechannel}.  In both cases, there is a asymptotic $T^{-2}$ behavior in the low temperature limit.  (\textbf{e})  Comparison of the dynamic viscosity, $\eta$, of the electron fluid in cadmium with $^3$He \cite{viscosity-He}, $^4$He \cite{He4}, water and olive oil \cite{oil}.
}
\label{fig.3}
\end{figure*}

 Expressing now  resistivity as $\rho (s, T) = \rho(0,s) + AT^{\alpha}$, where $\rho (0, s)= \rho (0, \infty) + as^{-1}$, the logarithmic derivative yields $\alpha=\frac{\partial \rm{ln}[\rho-\rho(0,s)]}{\partial {\rm{ln}}T}$. The result is shown in Fig.\ref{fig.2}a. One can see that the upturn in resistivity for $s\leq 64.8 \rm{~\mu m}$, leads to $\alpha \sim -2$ between 2 K and 5 K. The temperature and size dependence of the intrinsic electrical conductivity after subtraction of the size-dependent residual resistivity, $\Delta \sigma (T,s)= [\rho-\rho(0,s)]^{-1}$, is shown in Fig.\ref{fig.2}b.  Below $T$ = 5 K,  $\Delta \sigma \propto T^{2}$. This $T$-square dependence is the canonical hydrodynamic behavior.  A deviation from this behavior, visible at low temperature ($T<2.5~$K) and small thickness ($s <20~\mu$m) is a precursor of the ballistic regime. As discussed below, the hydrodynamic regime vanishes when the momentum conserving mean free path exceeds the sample thickness.   

\begin{table*}[ht!]
\centering
\renewcommand{\arraystretch}{1.4}
\begin{tabular}
{|m{4.7cm}<{\centering}|m{1.7cm}<{\centering}|m{1.8cm}<{\centering}|m{1.8cm}<{\centering}|m{1.7cm}<{\centering}|m{2.2cm}<{\centering}|m{1.6cm}<{\centering}|}  
\hline
System & $T_F$ (K) & $v_F$ (km/s)  & $\nu T^2$ (m$^2$s$^{-1}$K$^2$) &  $\tau_{mc} T^2$ (psK$^2$) & Temperature window (K) & Reference \\
\hline
\hline
$^3$He ($n$ = 1.63$\times 10^{22}~\rm{cm}^{-3}$) & 2 & 0.057   & $4.2\times 10^{-9}$ & 1.3 & $<$ 0.02 & \cite{Bertinat1974,Wheatley1975}\\
\hline
graphene ($n$ = 0.5$\times 10^{12}~\rm{cm}^{-2}$) & 1800 & 1000  & 970 & 3200 & 20--70 & \cite{Polini-viscouselectronliquid,graphitechannel} \\
\hline
Cd  ($n = p = 3.48\times 10^{22}~\rm{cm}^{-3}$) 
& $5.6$--$35000$ 
& $1600$ 
& $3.98$ 
& $6.2$ 
& 2--4 
& \cite{Subedi2024,Gall2016} \\
\hline
\end{tabular}
\caption{ \textbf{Comparison of three quantum fluids.} The Fermi temperature, the Fermi velocity and the prefactor of $T$-square viscosity in $^3$He, in graphene, and in cadmium. In the latter case, there are multiple pockets, lengths and energy scales. }
\label{Table_1}
\end{table*}

The hydrodynamic contribution to the electrical conductivity scales with the momentum-conserving collision time,  $\tau_{mc}$ following \cite{PhysRevLettnegative,Levitov2016}:

\begin{equation}
\label{delta-s}
\Delta \sigma=\frac{1}{3}\frac{ne^2}{m^*}\frac{s^2}{v_F^2\tau_{mc}}
\end{equation}

This expression is reminiscent of the Drude conductivity, with scattering time replaced by a time scale of $\frac{s^2}{v_F^2\tau_{mc}}$ and with $\ell_{mc} = \tau_{mc}v_F$. Fig.\ref{fig.2}c is a plot of $\tau_{mc}$ for five samples  derived from the experimental data and Eq. \ref{delta-s} using $v_F=1600$ km/s and $n=3.48 \times 10^{22}$ cm$^{-3}$. Also shown is $\tau_{mr}$ extracted from the temperature dependence of resistivity in the thickest sample (See Supplementary Note 2 for details \cite{SM}).

\begin{figure*}[ht!]
\includegraphics[width=1.0\linewidth]{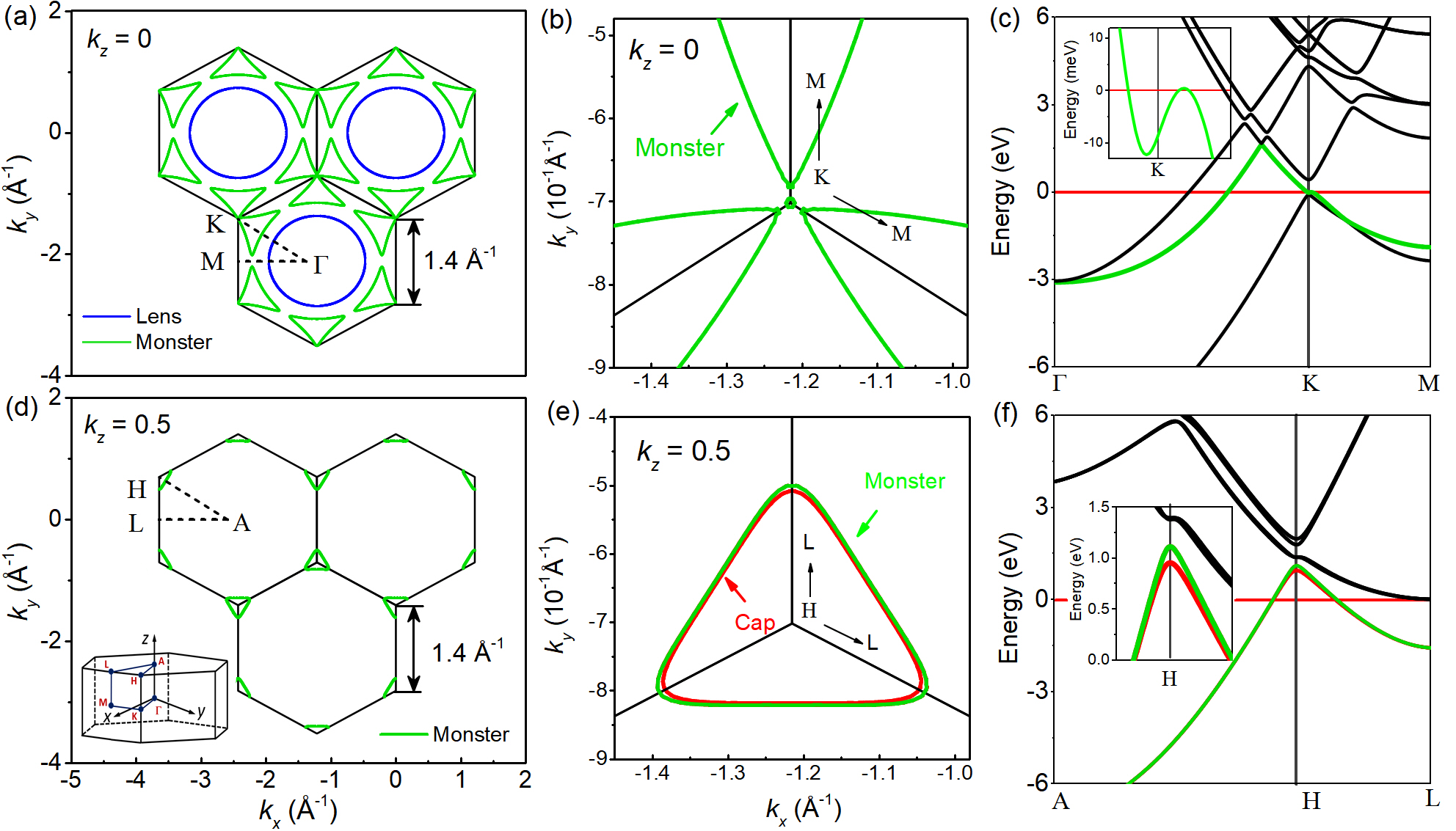} 
\centering
\caption{\textbf{Fermi surface cross section and band structure in HCP Cd.}   Cross section of the Fermi surface at $k_z = 0$, covering three Brillouin zones. The `monster' and `lens' contours are indicated by green and blue lines. (\textbf{b}) Zoom-in view of the `monster' hole pocket around the K point. (\textbf{c}) The band structure along the $\Gamma \rightarrow$ K $\rightarrow$ M direction. (\textbf{d}) The contours of the `monster' are specified at $k_z = 0.5$. (\textbf{e}) A zoomed-in view at $k_z = 0.5$ shows that the `monster' and `cap' hole-type pockets are close to each other but are not degenerate. (\textbf{f}) The band structure along the A$ \rightarrow$ H $\rightarrow$ L direction.
}
\label{fig.4}
\end{figure*}

As the thickness is decreased further, the hydrodynamic window closes. Data obtained on samples thinned below 10 $\mu$m are shown in Fig.\ref{fig.3}\textbf{a}.  The hydrodynamic upturn resistivity is gradually suppressed with decreasing $s$.  When $s <5.6 \mu$m and $T< 3.5$ K, $\rho \propto$ $T^{0}$: the resistivity becomes temperature-independent, uniquely depending on $s$ \cite{SM}, indicating a transition from the hydrodynamic to the ballistic regime. This is consistent with  Gurzhi's picture \cite{gurzhi1968,gurzhi-2,gurzhi-3}. The temperature dependence of the resistivity of the $s= 2$ $\mu$m sample is shown in Fig.\ref{fig.3}b. A modest resistivity upturn below 5 K abruptly stops at $\sim4$ K. At lower temperatures, resistivity becomes flat, with the extracted mean free path of the order of thickness. Fig.\ref{fig.3}c shows the ratio of $\frac{\rho ({\rm{5K}} , s)}{\rho ({\rm{2K}} , s)}$ as a function of $s^{-1}$. It becomes less than 1 in the hydrodynamic regime,  sandwiched between the diffusive and ballistic regimes as the size shrinks.

The  kinematic viscosity also proportional to $\tau_{mc}$:  $\nu = \frac{1}{4}v_F^2 \tau_{mc}$ \cite{viscosity-1,viscosity-2,viscous-4} and can be extracted from the experimental  $\Delta \sigma$.  Fig.\ref{fig.3}d  compares the extracted $\nu$ with what was reported in graphene \cite{graphitechannel}. In both cases, there is a $T^{-2}$ dependence at low temperature. Multiplying $\nu$ by the mass density leads to dynamic viscosity, $\eta$.  Fig.\ref{fig.3}e compares its amplitude and the temperature dependence with that of four liquids, two quantum (He isotopes) and two classical (water and olive oil). In our limited temperature range, the viscosity of electrons in Cd is larger than the universal lower bound observed in a variety of fluids \cite{Trachenko2021}.

Table \ref{Table_1} compares three liquids of fermions ($^3\text{He}$, graphene, and Cd). Their prefactor of $T$-square kinematic viscosity ($\nu T^2$)  differ by orders of magnitude. It is surprising to see that time scale of fermion-fermion collisions, in cadmium is almost as frequent as those in $^3\text{He}$. It is also worth comparing momentum-conserving  and momentum-relaxing collision rates. The amplitude of the prefactor of $T-$square resistivity $A$ in  bulk Cd yields  $\tau_{mr}T^2 \sim 2.4 \times 10^3 ~\text{ps}\cdot\text{K}^2$, three orders of magnitude larger than $\tau_{mc}T^2$. 
 
In a single-component isotropic Fermi liquid with $\mathcal{O}(1)$ Landau parameters, the fermion-fermion collision rate is set by the Fermi energy \cite{giuliani2005quantum,Dobbs,Vollhardt1990}:
\begin{equation}
\tau_{col.} \approx \frac{\hbar E_F}{(k_BT)^2}
\end{equation}

A Fermi energy of  $\approx 1$ eV, the order of magnitude for the large Fermi surface pockets in Cd,  would give a reasonable account of its $\tau_{mr}T^2$,  but not of its much shorter $\tau_{mc}T^2$. The latter requires the relevance of much smaller energy scales. \textit{Ab initio} calculations incorporating spin-orbit coupling (See supplementary note 3 \cite{SM}) finds an intricate electronic band structure (Fig. \ref{fig.4}) with multiple Fermi energies (Table \ref{tab:fermi}). Indeed, $\tau_{mc}T^2$ is roughly compatible with the lilliputian Fermi energy near the K-point. Its amplitude is at the edge of accuracy of these calculations.  The existence of $\approx$ 1 K energy scales is also inferred by the occurrence of magnetic breakdown  at very low fields \cite{Tsui1966,Datars1969,Visscher1970} in Cd. 

\begin{table}[htbp]
\renewcommand{\arraystretch}{1.4}
\centering
\caption{\textbf{Calculated Fermi energies of different pockets.} The bottom of the electron-like pocket known as `lens' at $\Gamma$ is taken as its Fermi energy. The tops of the hole-like `cap' and the hole-like `monster' at H define their Fermi energies. In addition, there are small energies at K-point: Petty-1 is associated with the `cap' and petty-2 with the `monster'. Along K-M, the lilliputian energy corresponds to a tiny local maximum (See the inset of Fig. \ref{fig.4}c).}
\label{tab:fermi}
\begin{tabular}{lc}
\toprule
Pocket & Fermi energy (eV) \\
\midrule
Lens        & \(-3.05\)           \\
Monster     & \(+1.62\)           \\
Cap         & \(+0.96\)           \\
Petty-1     & \(-8.8\times 10^{-2}\) \\
Petty-2     & \(-1.2\times 10^{-2}\) \\
Lilliputian & \(+4.8\times 10^{-4}\) \\
\bottomrule
\end{tabular}
\end{table}

The Fermi surface geometry is known to play a role in the robustness of the hydrodynamic regime \cite{Cook2019}.  In cadmium, as seen in Fig.\ref{fig.4}, the geometry is intricate. The three neighboring valleys (`monsters') almost touch at each K-point of the Brillouin zone.  The `monster' and the `cap' become almost degenerate near each H-point. These details may play a role in widening the large gap between $\tau_{mc}T^2$ and  $\tau_{mr}T^2$.

Exchange of phonons between electrons may play a role in amplifying momentum-conserving $e$-$e$ collisions \cite{jaoui2022formation,review-2,VoolImaging,LEVCHENKO2020168218}. Interestingly, Cd has a large e-ph collision cross section (See supplementary note 4 \cite{SM}).

In summary, thanks to thinning Cd crystals down to micrometric dimensions, we observed a canonical hydrodynamic transport regime sandwiched between the diffusive and the ballistic regimes. The frequency of momentum-conserving $e$-$e$ collisions  is set by an energy scale much smaller than the main Fermi energy. The computed band structure confirms the existence of a Lilliputian Fermi energy. A quantitative account of the rate of the fermion-fermion collisions from inter-valley and inter-pocket bottlenecks emerges as a task for future studies. 
 
We thank M. Feigel'man and A. Lucas for helpful discussions. This study, part of a Cai Yuanpei Franco-Chinese program (No. 51258NK), was supported by the National Science Foundation of China (Grant No. 12474043, 12004123 and 51861135104), the National Key Research and Development Program of China (Grant No. 2022YFA1403500), and the Fundamental Research Funds for the Central Universities (Grant no. 2019kfyXMBZ071).  X. L. was supported by The National Key Research and Development Program of China (Grant No.2023YFA1609600) and the National Science Foundation of China (Grant No. 12304065).

\clearpage

\begin{center}{\large\bf Supplementary Materials for ``Hydrodynamics of the viscous electron fluid in cadmium"}\\
\end{center}

\renewcommand{\thesection}{Supplementary\arabic{section}}
\renewcommand{\thetable}{S\arabic{table}}
\renewcommand{\thefigure}{S\arabic{figure}}
\renewcommand{\theequation}{S\arabic{equation}}

\setcounter{section}{0}
\setcounter{figure}{0}
\setcounter{table}{0}
\setcounter{equation}{0}

\section{Supplementary Note 1. Materials and Methods}

\textbf{Crystal growth.} High quality single crystals of Cd were grown by the vapor-phase transport method. The quartz tube with appropriate amount of Cd (99.9997\%) was evacuated and kept at the growth temperature for two weeks in a two-zone furnace. Shiny plate-like crystals were produced in a temperature range of 200 – 300$^{\circ}$C.  

\textbf{Sample fabrication.}  
 First, the crystallographic orientation of the bulk crystal was determined using X-ray diffraction. Subsequently, the crystal was securely affixed to a SEM stub in the desired orientation. The critical step was to employ a focused beam of Ga ions to meticulously carve a rectangular slab, a lamella, from the crystal. This process was carried out in three stages.

\begin{table}[ht]   
\begin{center} 
\label{table:S1} 
\begin{tabular}{|m{1.4cm}<{\centering}|m{2.9cm}<{\centering}|m{1.6cm}<{\centering}|m{1.6cm}<{\centering}|}   
\hline   
\textbf{Sample} & \textbf{$= w\times t$ ($\rm{\mu m^2}$)} & \textbf{s ($\rm{\mu m}$)} & \textbf{$L$ ($\rm{\mu m}$)} \\   
\hline   $\#$1-1 & 4 $\times$ 33.6  & 11.6 & 208 \\ 
\hline   $\#$1-2 & 8 $\times$ 33.6  & 16.4  & 208 \\  
\hline   $\#$1-3 & 17 $\times$ 33.6  & 23.9 & 208 \\  
\hline   $\#$1-4 & 50 $\times$ 33.6  & 41.0  & 208\\ 
\hline   $\#$1-5 & 125 $\times$ 33.6  & 64.8 & 208\\
\hline   $\#$1-6 & 130.5 $\times$ 100.3 & 114.4 & 1000\\
\hline   $\#$2-1 & 0.67 $\times$ 6.21  & 2.02 & 42 \\
\hline   $\#$2-2 &  1.21 $\times$ 6.21 & 2.74 & 42 \\
\hline   $\#$2-3 &  3.0 $\times$ 6.21  & 4.3 & 42 \\
\hline   $\#$2-4 &  5.0 $\times$ 6.21 & 5.6 & 42 \\
\hline   $\#$3-1 &  12 $\times$ 6.8  & 9.0 & 26.7 \\
\hline   $\#$3-2 &  25 $\times$ 6.8  & 13.0 & 26.7 \\ 
\hline 
\end{tabular}   
\end{center}
\caption{\textbf{Sample dimensions.} Cadmium crystals used in this study were all oriented along the [0001] crystallographic axis. $L$, $t$  and $w$ are, respectively, the length, the thickness and the width of each sample. $s$ $=$ $\sqrt{w \times t}$ represents the average diameter of the conducting cross-section. }
\end{table}

Initially, a high current of 50 nA was used to create two gaps within the crystal. These gaps were separated by a section of the crystal, approximately $\sim$25 $\mu$m thick and 150 $\mu$m long. Following this step, a smaller current of 15 nA was applied to smoothen the crystal section, resulting in moderately flat sidewalls. 
The lamella remained attached to the parent crystal through two beams at its top.

Afterwards, with a precise 7 nA current both sides of the lamella were fine cut, ensuring parallel sidewalls and a high level of smoothness. The obtained flakes were then carefully transferred to an alumina substrate. It is crucial to note that a very thin layer of ``Araldite'' epoxy adhesive must be applied to the substrate beforehand. Once positioned on top of the epoxy adhesive, capillary forces naturally shaped the epoxy around the lamella, extending smoothly to each of its top surfaces without covering them.

Once the  epoxy adhesive has dried, the substrate was taken to a sputtering machine, depositing a 300 nm layer of Au onto the lamella. To further refine the device, the FIB system was utilized. Initially, the Au layer covering the active part of the device was removed using an acceleration voltage of 5 kV and an ion current of 2 nA. Subsequently, the overall device shape, including the contact positions, was precisely cut out at 0.3 nA and 30 kV. Finally, the Au layer away from the device was severed, ensuring a current flow exclusively through the device. With these steps completed, the device becomes ready for measurement.

The samples used in this study are presented in table. S1. We fabricated three samples ($\#1$, $\#2$, $\#3$) of different thicknesses. For each, the thickness was held constant while the width $w$ was progressively reduced to vary the effective cross-sectional area $s$.


\section{Supplementary Note 2. Variation of residual resistivity with boundary scattering}

In the Fuchs-Sondheimer (FS) transport theory \cite{Zhou2018}, resistivity follows  the following dependence on thickness:
\begin{equation}
\rho= \rho_0 \left[1+ \frac{3\ell_{mr} (1-p)}{8s}\right]
\end{equation}

Here, $p$ is the specularity parameter and 
\begin{equation}
\rho_0^{-1}=\frac{2}{3\pi} \frac{e^2}{h}\ell_{mr}(k_{F,h}^2+k_{F,e}^2)
\end{equation}
$k_{F,h}$ and $k_{F,e}$ are Fermi wave-vectors for electrons and hole pockets. This implies such a relation between $a$, $k_F$ and $p$:

\begin{equation}
a= \frac{9\pi}{32}\frac{h}{e^2}\frac{1-p}{k_{F,h}^2+k_{F,e}^2}
\end{equation}

Experimentally, we find  $a=8.8 \pm 0.3 \times 10^{-16} ~\rm{\Omega \cdot m^2}$. Given that in Cd,  $ k_{F,e}$ is between 2.5 and 7.5 $~\rm{nm^{-1}}$ \cite{Subedi2024}, our measured $a$ indicates $0.2<p< 0.6$. It is common to find a specularity parameter between 0.5 and 0.9 in elemental metals \cite{Tsoi1999}.





\section{Supplementary Note 3. Fermi surface DFT theoretical Calculation}
We performed the electronic structure calculations of cadmium both with and without taking into account the spin-orbit interaction.  The calculations were done within the local density approximation (LDA) using the full potential {\sc wien2k} package \cite{wien2k}.  A muffin-tin radius $R$ of 2.5 a.u.\ was used for Cd, and the plane-wave cutoff $K_{\textrm{max}}$ was set using $RK_{\textrm{max}} = 7$. A $172 \times 172 \times 86$ $k$-point grid was used for the Brillouin zone integration. The Fermi surface was obtained on a $k$-point grid of  $180 \times 180 \times 90$.  

\begin{figure*}[ht]
\includegraphics[width=0.8\linewidth]{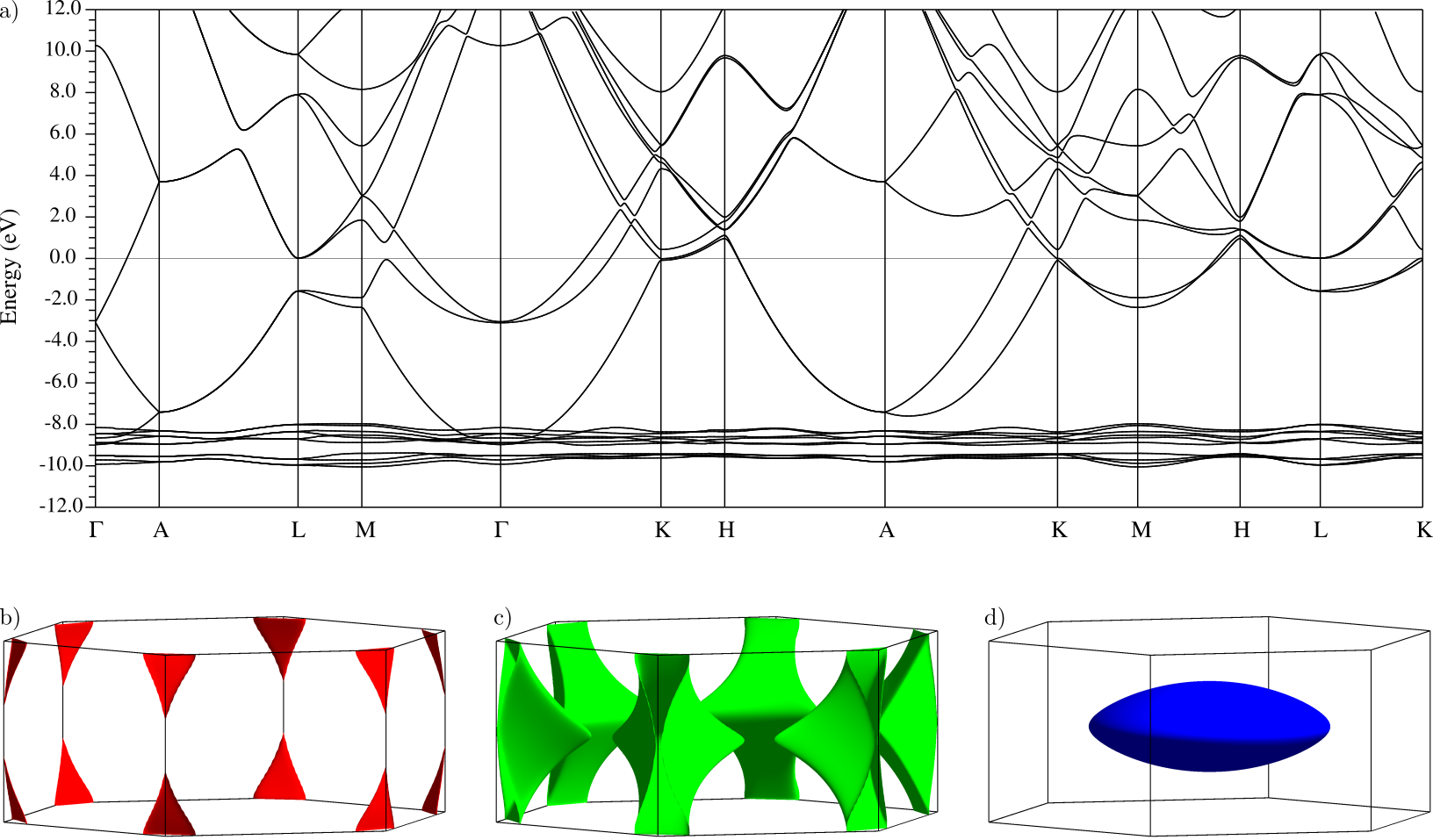} 
\centering
\caption{\textbf{Electronic structure of HCP Cd including spin-orbit interaction.} a) The band structure features a pair of bands that disperse linearly along the out-of-plane direction $\Gamma$-$A$ and quadratically along the in-plane direction $\Gamma$-$M$ and $\Gamma$-$K$. Individual Fermi sheets consisting of b) cap, c) monster, and d) lens sheets.  Note that these are similar to the results shown in Ref.~\cite{Subedi2024} that does not take into account the spin-orbit interaction.}
\label{fig.S1}
\end{figure*}

\begin{figure*}[ht]
\includegraphics[width=0.9\linewidth]{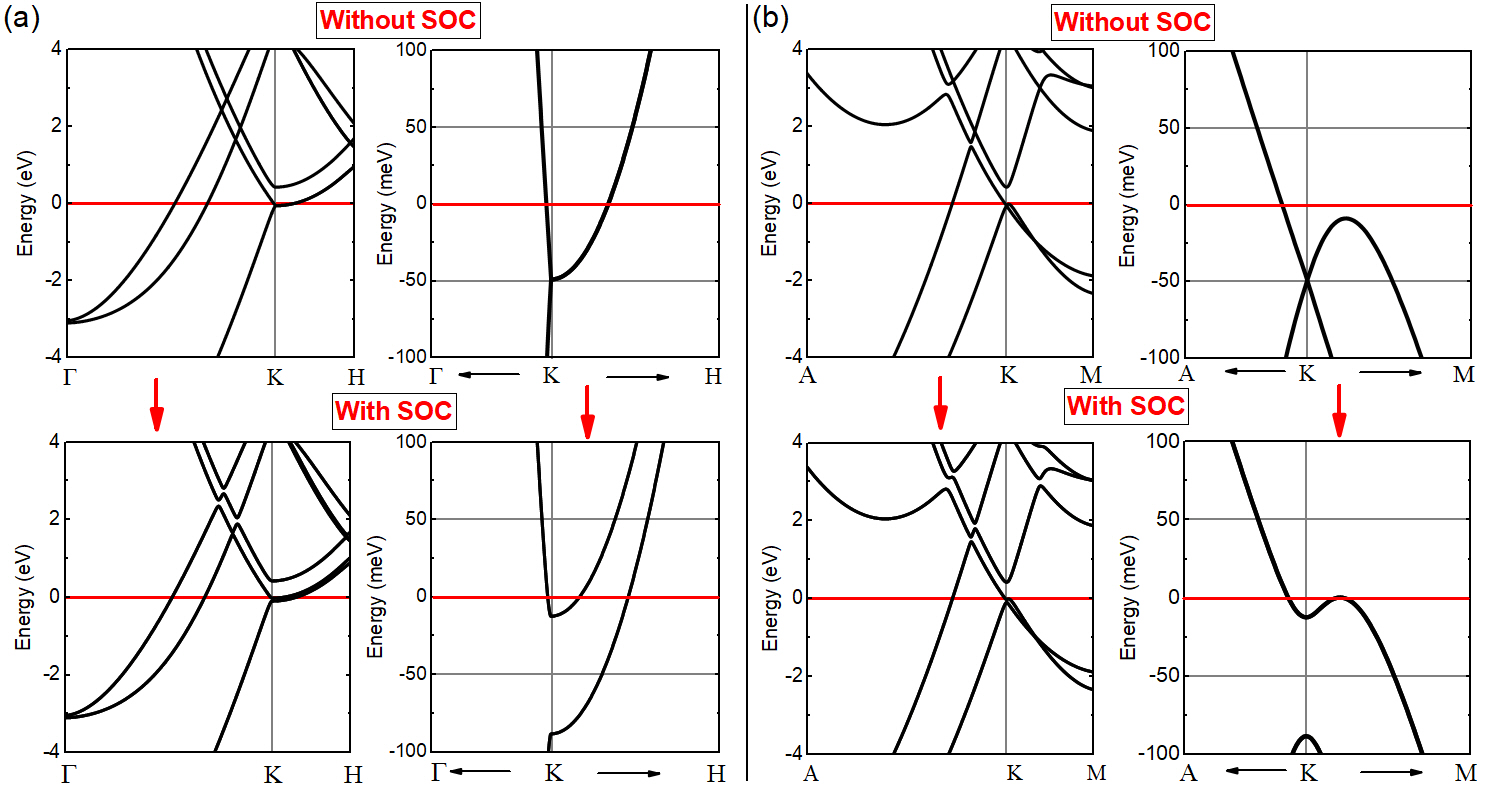} 
\centering
\caption{\textbf{Band structure near Fermi level.} \textbf{(a)} The band structure along $\Gamma \rightarrow K \rightarrow H$ direction. A very small electronic pocket can be observed near the K point. Taking spin-orbit coupling into account makes the pocket significantly smaller, but it remains present. \textbf{(b)} The band structure along $A \rightarrow K \rightarrow M$ direction. A very small electronic pocket can be observed near the K point. The pocket is still there, even when spin-orbit coupling is taken into account.   }
\label{fig.S2}
\end{figure*}

The three-dimensional Fermi surface, its planar cuts, and the band structure shown in Figs.\textbf{1a} and \textbf{4} include spin-orbit interaction. In Fig.~\ref{fig.S1} we show the band structure along an extensive high-symmetry path within the first Brillouin zone and the individual Fermi sheets, which also take into account the spin-orbit interaction.  We note that the electronic structure calculations in a recent work by some of us do not take into account the spin-orbit interaction \cite{Subedi2024}. However, the effect of spin-orbit interaction is negligible except at certain band degeneracies at high-symmetry points.  As can be seen in Fig.~\ref{fig.S2}, the degeneracy at $K$ near the Fermi level is lifted by $\sim$100 meV.

\begin{figure*}[ht]
\includegraphics[width=0.9\linewidth]{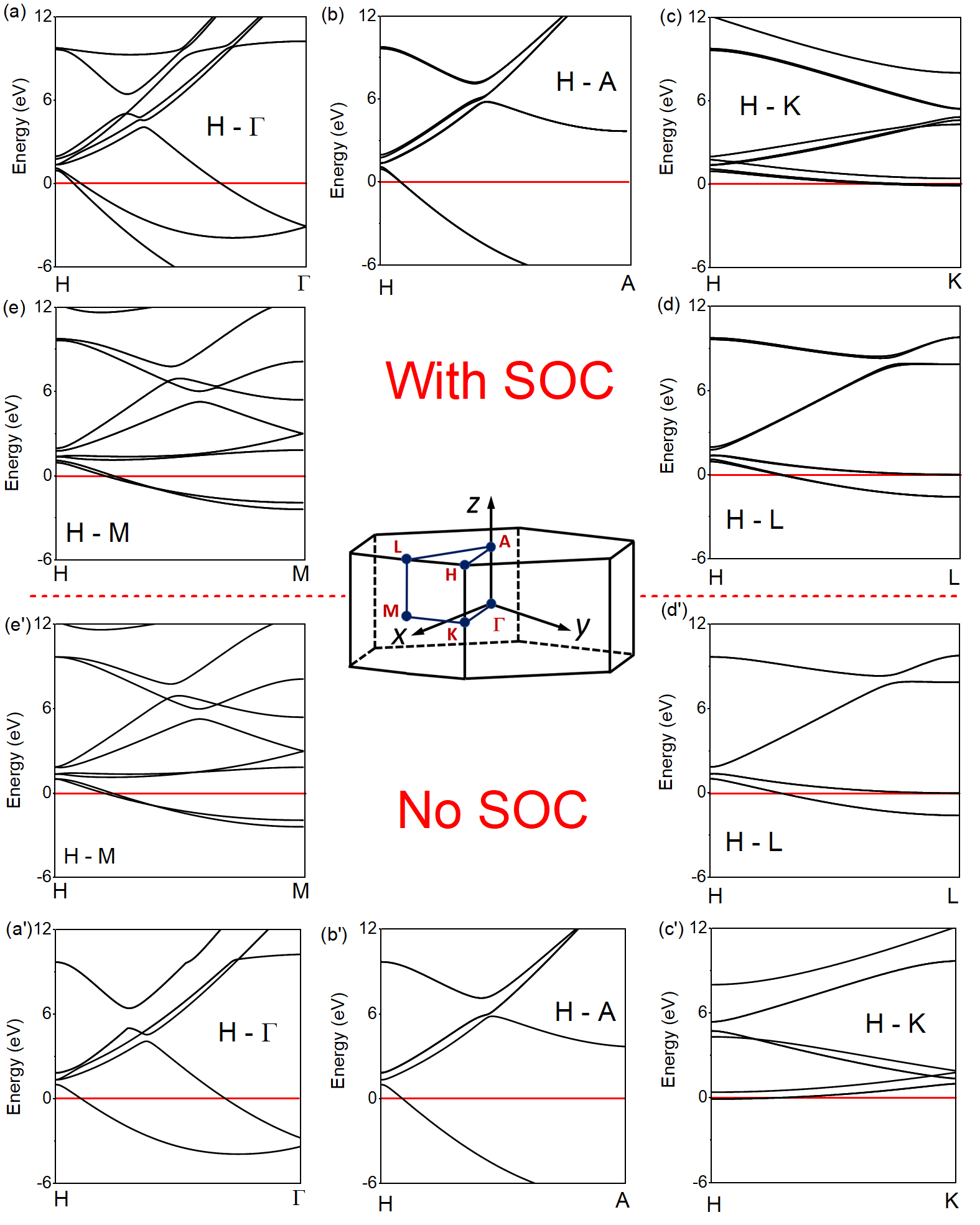} 
\centering
\caption{\textbf{The Fermi surface in cadmium around H point.}  }
\label{fig.S3}
\end{figure*}

\begin{figure*}[ht]
\includegraphics[width=0.9\linewidth]{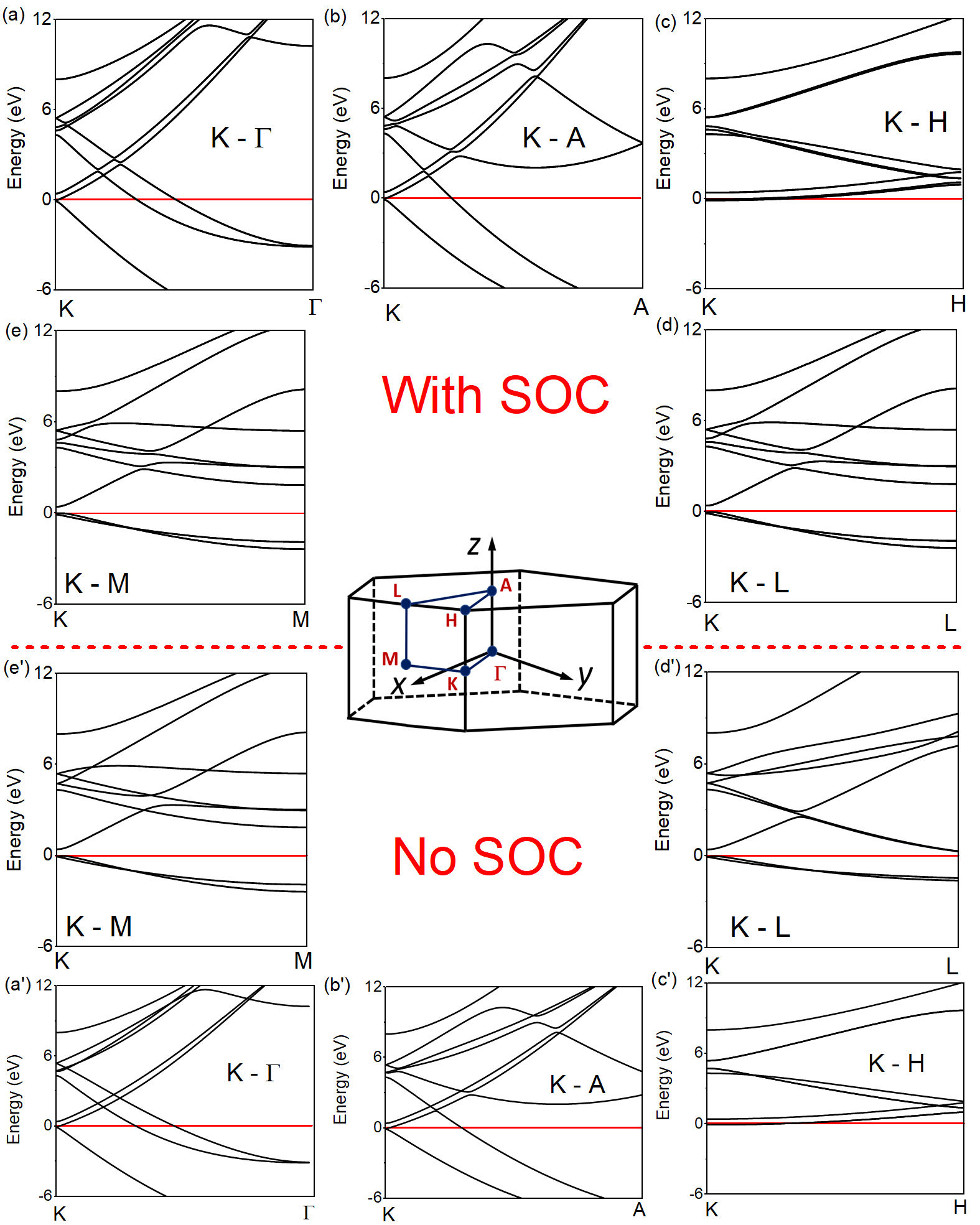} 
\centering
\caption{\textbf{The Fermi surface in cadmium around K point.}  }
\label{fig.S4}
\end{figure*}

\begin{figure*}[ht]
\includegraphics[width=0.95\linewidth]{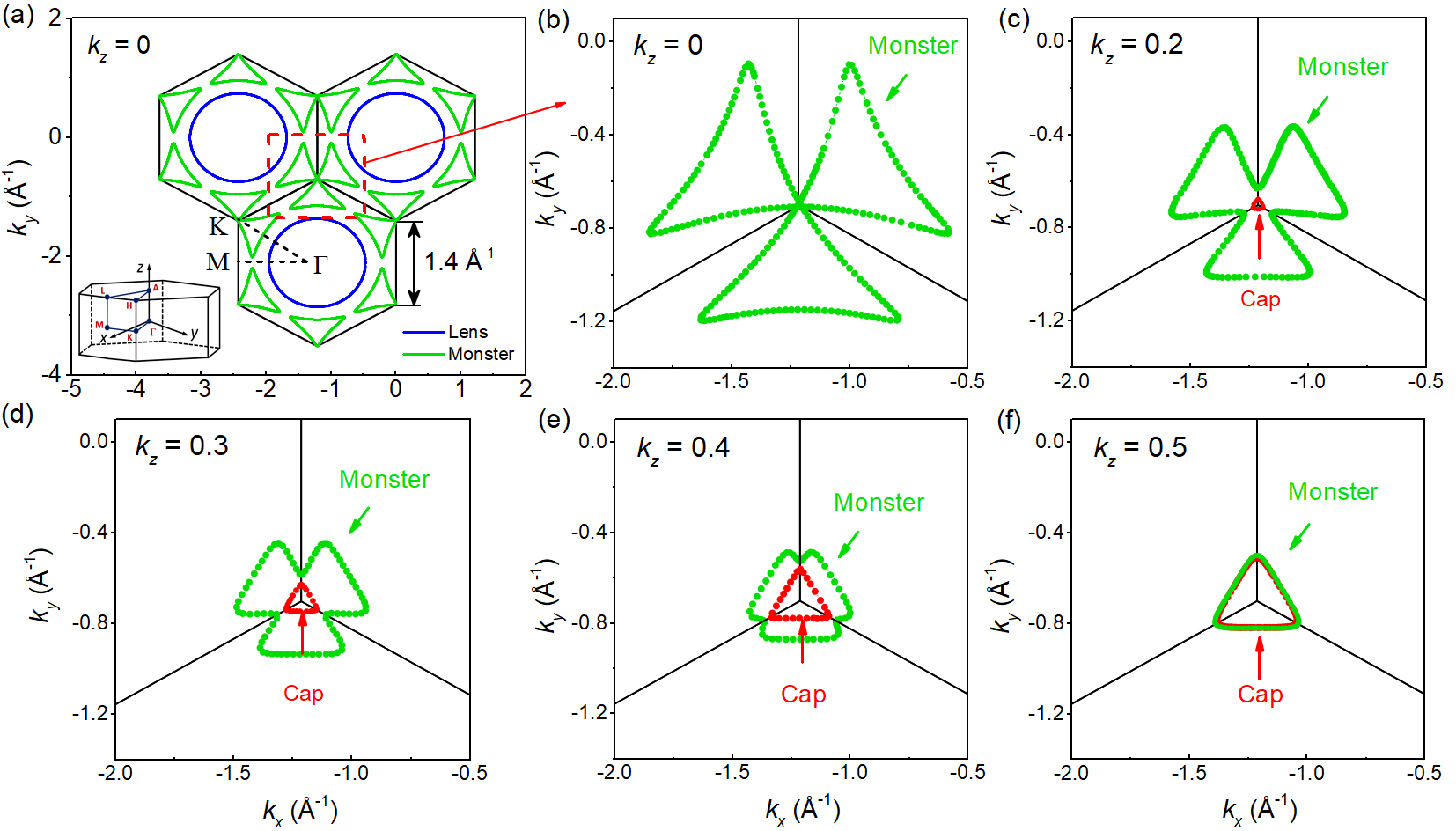} 
\centering
\caption{\textbf{The Fermi surface cross section as a function of $k_z$ in cadmium.}  }
\label{fig.S5}
\end{figure*}

\section{Supplementary Note 4. The prefactors of electron-phonon and electron-electron scattering terms in cadmium}

\begin{figure*}[ht]
\includegraphics[width=0.5\linewidth]{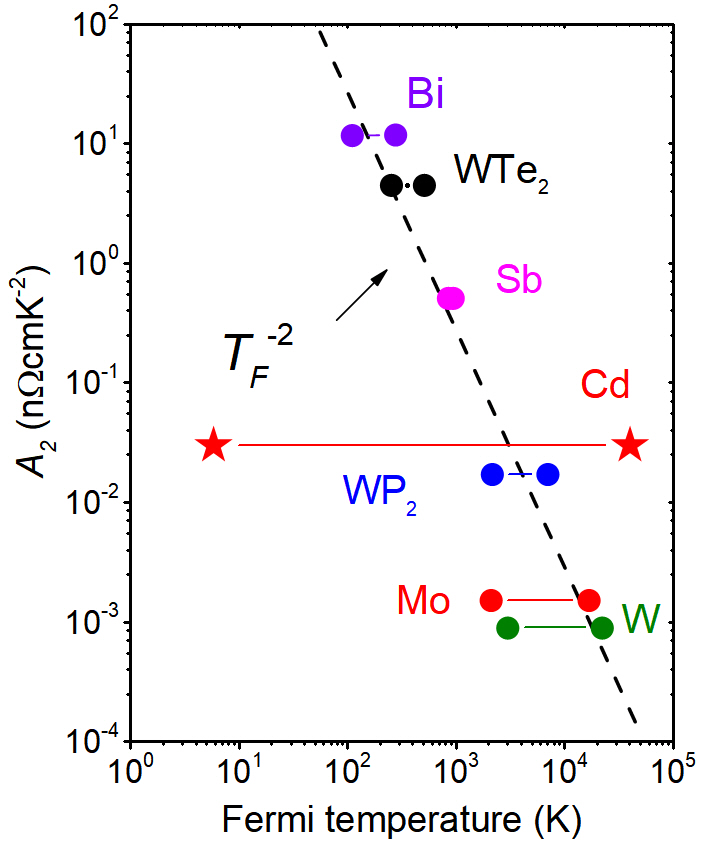} 
\centering
\caption{\textbf{The prefactor of $T^2$ resistivity in cadmium compared to other elemental metals.}The prefactor of $T$-square electrical resistivity, $A_2$, as a function of the Fermi energy in different semimetals.}
\label{fig.S6}
\end{figure*}

\begin{figure*}[ht]
\includegraphics[width=0.8\linewidth]{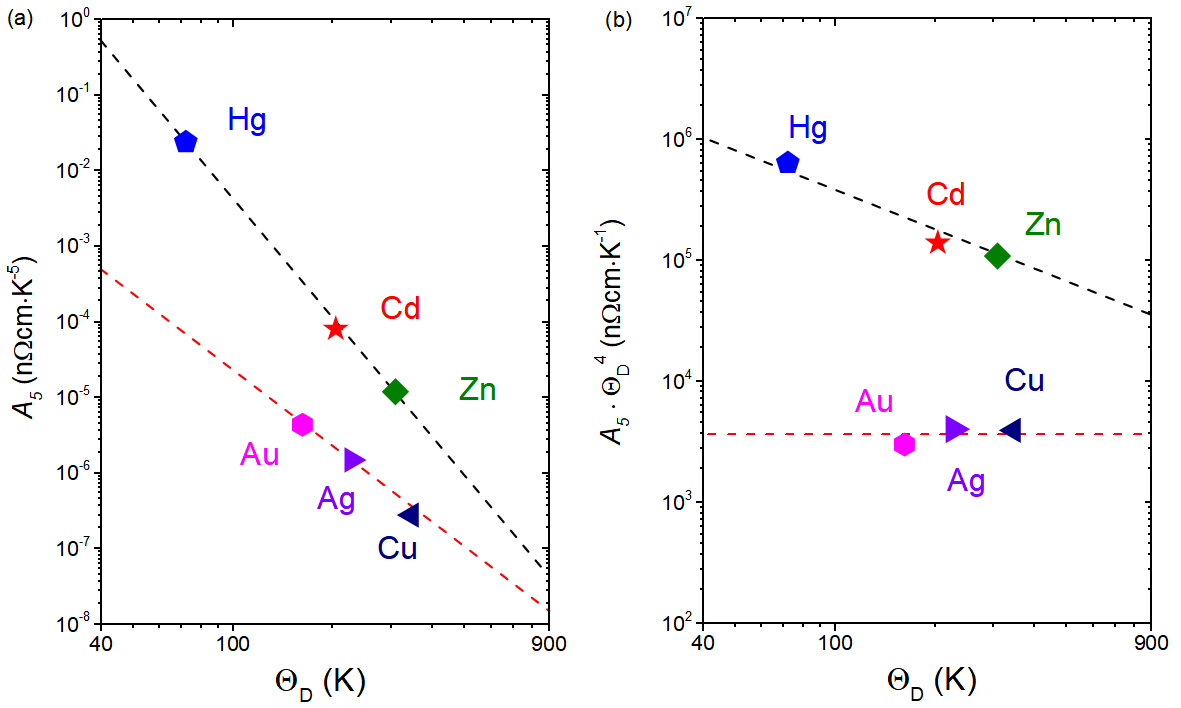} 
\centering
\caption{\textbf{The prefactor of $T^5$ resitivity in cadmium compared to other elemental metals.} (\textbf{a}) Amplitude of $A_5$ \cite{Matula1979,Desai1984,van1964,ryan1980resistivity} as a function of Debye temperature \cite{Martin1973,Seidel1958,Rajdev1962,RYAN1980158} in six metals in two neighboring columns of the periodic table. Dashed lines are guides to the eye. Cadmium has the largest $A_5$, because it has a low Debye temperature and a large electron-phonon cross section. (\textbf{b}) The perfactor $A_5$ scaled by $\Theta_D^4$.  }
\label{fig.S7}
\end{figure*}

The prefactor of $T-$square electrical resistivity, $A_2$, as a function of the Fermi temperature in different semi-metals as shown in Fig.\ref{fig.S6}. The dashed line represents a $T_F^{-2}$ slope. For each case, the two symbols indicate the largest and the lowest Fermi energy.  In  cadmium, smallest Fermi energy is 15meV (See Fig.\textbf{4c}).

Figure \ref{fig.S7}\textbf{a} is a plot of the amplitude of $A_5$, the prefactor of $T^5$ electrical resistivity, in six elemental metals, belonging to two distinct families. As expected \cite{ziman2001electrons}, the amplitude of $A_5$ correlates with the Debye temperature, $\Theta_D$ in each family. Compared to Zn, Cd has a lower Debye temperature. More importantly, the electron-phonon scattering cross section is larger in the Group 12 elements (Cd and Zn, with a HCP structure) than in the noble metals (Cu, Ag and Au, with a FCC structure).  Theory \cite{review-2,LEVCHENKO2020168218} and experiment \cite{jaoui2022formation} suggest that normal electron-electron collisions can occur by  exchange of acoustic phonons. A large electron-phonon cross section may therefore play a role in the robustness of electron hydrodynamics in cadmium.

\section{Supplementary Note 5. Subtracted the term of T-square from the longitudinal conductivity}

\begin{figure*}[ht!]
\includegraphics[width=0.8\linewidth]{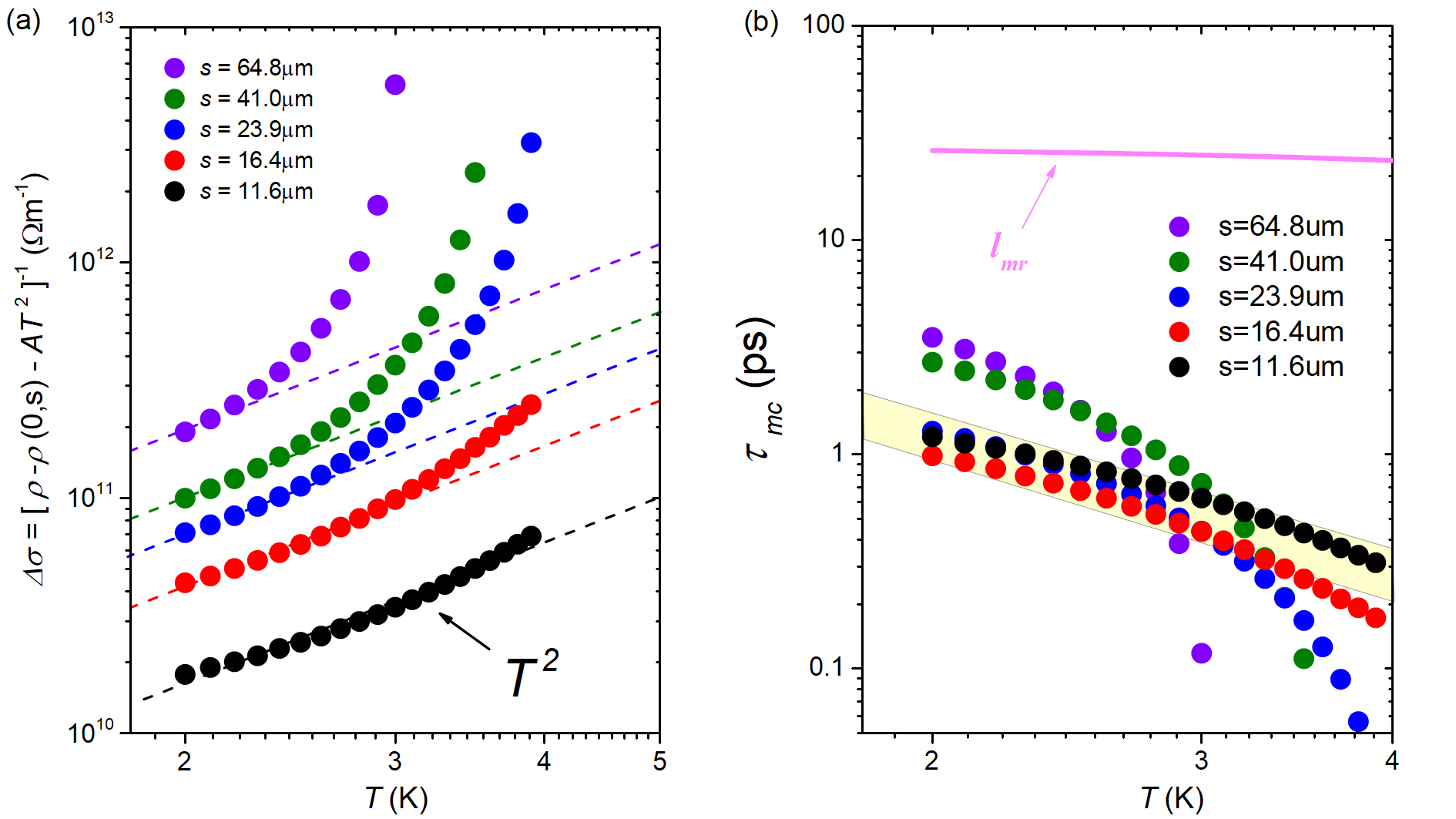} 
\centering
\caption{\textbf{Extracting the hydrodynamic contribution to electrical conductance.} (\textbf{a}) Temperature dependence of $\Delta \sigma =1/[\rho-\rho(0,s)-AT^2]$. (\textbf{b}) Evolution of $\tau_{mc}$ for different samples. }
\label{fig.S8}
\end{figure*}

In the main text, we focus exclusively on the impact of sample boundary scattering on longitudinal conductivity at low temperatures, $\Delta \sigma  = [\rho-\rho(0,s)]^{-1}$, as shown in Fig.2\textbf{b}. However, in practice, momentum relaxation between electrons persists and must be accounted for. Previous reports\cite{jaoui2021thermal,jaoui2022formation} indicate that component $A$ increases with sample size. However, the resistivity upturn at low temperatures impedes a conclusive analysis of this trend. Here, we treat it as a constant, $A = $ 0.03 n$\Omega$cm$\cdot$ K$^{-2}$, independent of sample size. As a result, as shown in Fig.\ref{fig.S8}, the term of $T$-square contributes little to the longitudinal conductivity $\Delta \sigma$ and can be neglected at low temperatures.

\end{document}